\documentclass[11pt]{article}
\usepackage{latexsym}
\usepackage{amssymb}
\usepackage{amsmath} 

\newtheorem{theorem}{Theorem}

\newenvironment{proof}{\noindent{\bf Proof:} }{\hfill$\Box$\medskip}

\newcommand{\strong}{{\sf strong}}
\newcommand{\accept}{{\sf accept}}
\newcommand{\weak}{{\sf weak}}

\begin{document}

\title{Nondeterministic One-Tape Off-Line Turing Machines and Their Time Complexity\,\footnotemark
\footnotetext[1]{Partially
supported by MIUR under the project PRIN ``Aspetti matematici e applicazioni emergenti degli automi e dei linguaggi formali: metodi probabilistici e combinatori in ambito di linguaggi formali''.}
}

\author{%
Giovanni Pighizzini
\mbox{}\\
{\normalsize Dipartimento di
    Informatica e Comunicazione}\\
{\normalsize Universit\`{a} degli Studi di Milano}\\
{\normalsize via Comelico 39, 20135 Milano, Italy}\\
{\normalsize\tt pighizzini@dico.unimi.it}%
}%
\date{}%
\maketitle\thispagestyle{empty}%

\begin{abstract}
In this paper we consider the time and the crossing sequence
complexities of one-tape off-line Turing machines.
We show that the running time of each nondeterministic machine accepting 
a nonregular language must grow at least as $n\log n$, in the case all accepting
computations are considered (\accept\ measure). 
We also prove that the maximal length of the crossing sequences used in 
accepting computations must grow at least as $\log n$.
On the other hand, it is known that if the time is measured considering,
for each accepted string, only the faster 
accepting computation (\weak\ measure), then there exist
nonregular languages accepted in linear time. We prove that under this measure,
each accepting computation should exhibit a crossing sequence of length at least
$\log\log n$.
We also present efficient implementations of algorithms
accepting some unary nonregular languages.

\end{abstract}

\section{Introduction}

One of the main problems in the design of computer algorithms and in their
implementation is that of producing efficient programs, under the restrictions
given by the available resources.

For example, up to the first '80 years, when the central memories of
the computers were very small and the operating systems of personal
computers did not provide virtual memory functionalities, one of the
critical topics in the implementation of applications using significative
amounts of data was that of choosing data structures requiring only
a small amount of extra memory, besides the memory needed by the
actual data represented. Many times, computer programmers, in order to
cope with a restricted space availability, were forced to choose
data structures not efficient from the point of view of the time used by the
algorithms manipulating them.

In the last two decades, we assisted to a continue and enormous increasing
of the capacity of memory chips, together with a reduction of their costs.
As an effect, now we can write very large computer programs, implementing
sophisticated algorithms, and, at the same time, computer programs
can use all this memory capacity to efficiently manipulate huge amount
of data.\footnote{We should point out that, many times, it seems that
some computer programs use now an amount of memory which appears to be
excessive with respect to their functionalities. 
For example, 20 years ago we already used word processors implementing 
almost all the functionalities that the most common word processors  
today available offer. However, the memory requirements now are tremendous.
(Sometimes, I need to use a very common word processor whose current version 
requires an amount of RAM which is 1000 times larger than the amount 
required by the version of the same program I was using in 1989.)
It should be interesting to know how certain computer programs are able to
use so huge amounts of memory.}
However, in many situations, the space is still a critical resource.
For example, the memories of portable devices, like mobile phones, 
usually are very limited; also
the programs for embedded systems have strong memory restrictions.
This is the main reason for which the designers of the Java programming
language decided to provide, among the primitive data types representing
integer numbers, the types {\tt byte} (8 bits),
{\tt short} (16 bits) for the representation of ``small'' integers, 
besides the ``standard'' types {\tt int}
(32 bits) and {\tt long} (64 bits) \cite{JAVA}.

Thus, the investigation of computing with restricted resources is
always an interesting and important topic.
From a theoretical point of view, this subject has been considered
by studying and comparing the power of machine models using certain
amounts of resources. A special area of interest is the border 
between finite state devices (described by finite automata) and
more powerful devices.

Since early '60, many researches have been done in order to 
discover the minimal amount of computational resources needed by a 
machine model in order to be more powerful than finite state devices.
This kind of investigation can be formalized in the realm of
formal languages, by studying the minimal amount of resources
needed to recognize nonregular languages.

Among these researches, the most famous are those ones concerning 
space and time lower bounds.

The investigation of the space requirements for the recognition
of nonregular languages started with the works of Hartmanis, Lewis and
Stearns~\cite{LSH65,SHL65}, and Hopcroft and Ullman~\cite{HU69}.
(For surveys see, e.g.,~\cite{Mi92} and~\cite{Me08}. The last paper
presents also some new and recent results.)
The first works concerning time requirements have been done
by~Trakhtenbrot~\cite{Tr64}, Hennie~\cite{He65}, 
and Hartmanis~\cite{Ha68}. We now discuss more in details the time resource,
because it is the main subject of this paper.

In his seminal paper, Hennie considered a very restricted computational
model: the one-tape off-line Turing machine. This machine has just
one semi-infinite tape that at the beginning of the computation contains 
the input string (for this reason it is called ``off-line''). The content
of the tape can be modified during the computation. The original
model was deterministic.
The time is measured by taking into account the time used by the
machine on all input strings.
Hennie proved that if a machine of this kind works in linear time
then the accepted language is regular. Hence, the capability of
storing data on the tape is useless under this time restriction.
A stronger bound on the time was independently obtained by
Trakhtenbrot~\cite{Tr64} and by Hartmanis~\cite{Ha68}: they proved
that in order to recognize nonregular languages the time must grow
at least as $n\log n$.

A similar investigation has been done, later, for the
nondeterministic version of one-tape off-line machines.
In 1986, Wagner and Wechsung~\cite{WW86} provided a counterexample
showing that the $n\log n$ time lower bound cannot hold in the nondeterministic
case. Subsequently, in 1991, Michel~\cite{Mi91} proved the existence
of NP-complete languages accepted in linear time.\footnote{It is trivial to
observe that there are regular languages requiring linear time.} 
However, these results 
were obtained by measuring the time in an ``optimistic'' way, namely
by considering, among all accepting computations on a string,
the faster one (this measure is called \emph{\weak}).
On the other hand, in 2004, Tadaki, Yamakami, and Lin~\cite{TYL04}
proved that, by taking into account, for each input string, all the computations
(this measure is called \emph{\strong}), the $n\log n$ lower bound
holds even for the nondeterministic case.

In this paper, we further refine these results. Besides the \strong\
and the \weak\ measures, we consider an intermediate measure called
\emph{\accept}. This measure considers the costs of all accepting
computations. (It has been already used in the
literature, mainly for the space complexity,
and differences with other measures have been shown, see~\cite{Me08}.)
First, we extend the $n\log n$ lower bound for nonregular language
acceptance in the nondeterministic case, from the \strong\ measure
to the \accept\ measure. Concerning the \weak\ measure, we are able to prove
that in order to accept a nonregular language, some tape cells
must be visited at least $\log\log n$ times, even if the language
is accepted in linear time. This will be proved using the concept
of \emph{crossing sequence}.\footnote{Two tables summarizing 
the already known lower bounds and the new results proved in this paper 
can be found in the last section.}

We will also present some examples of languages accepted by these
machines, obtaining fast implementations of some recognition
algorithms. All these examples are \emph{unary} of \emph{tally}
languages, namely languages defined over a one-letter alphabet.
So, the only information that the machine can use to accept
or reject an input string is its length.
We think that the techniques used to implement these algorithms
are interesting examples of programming with very restricted
resources. Furthermore, different implementations of the same
recognition algorithms, using restricted amounts of space,
have been presented in the literature.

\section{Preliminary notions and definitions}
\label{sec:prel}

As usual, we denote by $\Sigma^*$ the set of all strings over a finite 
alphabet $\Sigma$, by $\epsilon$ the empty string, and by $|x|$ the length
of a string $x\in\Sigma^*$.

A \emph{language} $L$ is a subset of $\Sigma^*$. $L$ is said to be a
\emph{unary} or \emph{tally language} if $\Sigma$ contains only one
element. In this case we stipulate $\Sigma=\{a\}$.

Given a language $L$ and an integer $n$, we denote by $L^{\leq n}$ the set
of strings of length at most $n$ belonging to $L$.

\smallbreak

In the paper we will consider \emph{one-tape off-line Turing
machines}~\cite{He65}. Besides a finite state control, these devices 
are equipped with a semi-infinite tape.
At the beginning of the computation the input string is written 
on the tape starting from the
leftmost tape square, while the remaining squares contain the blank symbol.
At each step of the computation, the machine writes a symbol on the
currently scanned square of the tape (possibly changing its content),
and moves its head to the left, to the right, or keeps it stationary,
according to the transition function. The machine never writes the blank
symbol. Hence, it is able to work and to modify
the part of the tape that, at the beginning, was containing the input string, as
the part of the tape to the right of the input string. However, after a
square has been visited, it will never contain the blank symbol.
Special states are designed as accepting and rejecting states. We assume
that in these states the computation stops.

To study the behavior of this kind of machines, 
it is useful the notion of \emph{crossing sequences}, that
we now recall~\cite{He65}.
Let us consider a deterministic or nondeterministic one-tape off-line
Turing machine $M$. Given a computation $\cal C$ of $M$ and a boundary
$b$ between two squares of the tape, the crossing sequence of $\cal C$
at $b$ is the sequence of the states $(q_1,q_2,\ldots,q_k)$, where
$q_i$, $i=1,\ldots,k$, is the state of $M$ when in the computation 
$\cal C$ the head crosses for the $i$th time the boundary $b$.
Note that for odd values of $i$ the state $q_i$ corresponds to a move
of the head from left to right, while for even values of $i$ to a move in the
opposite direction.
Hence, if the length $k$ of the crossing sequence $c$ is odd then the 
computation $\cal C$ must end with the
head scanning a square to the right of the boundary $b$,
while if $k$ is even then $\cal C$ must end with the head
scanning a square to the left of $b$.
A nonending computation could have crossing sequences of infinite length.
However, all the crossing sequences we will have to consider in our proofs
are finite.

We also point out that given two finite sequences $c',c''$ of states,
it is possible to verify whether or not they are ``compatible'' with respect
to a tape symbol $a$, namely, if $c'$ and $c''$ can be, respectively, 
the crossing sequence at the left and at the right boundary of a tape square
initially containing $a$. This test can
be done ``locally'', i.e., without knowing the rest of the tape content.
We now sketch a nondeterministic procedure performing
this test.

Let $c'=(q'_1,\ldots,q'_{k'})$, $c''=(q''_1,\ldots,q''_{k''})$ be
the two crossing sequences.
The head of the machine will reach the tape square under consideration 
for the first time with a move from the square to its left side.
Hence, the state during the first visit of the cell is $q'_1$.
The next move depends on $q'_1$ and on the original content $a$ of 
the square under consideration.
The procedure selects, in a nondeterministic way, one transition among the 
finitely many possible.
\begin{itemize}
\item If the selected transition moves the head to the left, in the 
  state $q'_2$, writing a symbol
  $b$ on the tape square, then the next visit to the tape square should
  be from the left side, in the state $q'_3$.
  Hence, the procedure continues in a similar way, 
  after replacing $c'$ with the shorter sequence
  $(q'_3,\ldots,q'_{k'})$, and by considering $b$ as content of the tape
  square.
\item If the selected transition moves to the right, 
  in the state $q''_1$, writing a symbol
  $b$ on the tape square, then the next visit to the tape square should
  be from the right side, in the state $q''_2$.
  In this case the verification procedure continues expecting
  the next move from the right side.  The crossing sequences $c'$ and
  $c''$  are replaced, respectively, by the shorter sequences
  $(q'_2,\ldots,q'_{k'})$ and $(q''_2,\ldots,q''_{k''})$. The square
  content considered will be the new symbol $b$.
  Hence,  at the next step, a transition
  from the state $q''_2$ with the symbol $b$ should be considered. 
\item If the transition does not move the input head (stationary move)
  a new move from the reached configuration is selected and the
  process is repeated (after updating the symbol on the square)
  until to select either a move to the left or 
  to the right, or to reach a configuration where the computation stops.
  Notice that if a finite sequence of stationary moves reaches a same 
  configuration twice, then there exists a shorter
  sequence of stationary moves ending as the original sequence. 
  The maximal number of stationary moves, before a repetition,
  is given by the product of the number of the states by the number of
  the possible symbols. Hence, if the procedure exceeds this number, it can
  stop and reject.
\item In the other cases the procedure stops and reject.
\end{itemize}
The procedure accepts if it is able to end with the two crossing sequences
empty. (Many technical details have been omitted.
For a more extended discussion about this topic, in the context of 
the reduction of two-way automata to one-way ones
we address the reader to~\cite{HU79}.)

\smallbreak

Given four strings $u,v,u',v'\in\Sigma^*$, and two computations
$\cal C$ and ${\cal C}'$ on inputs $uv$ and $u'v'$, respectively, suppose
that the crossing sequence $c$ of $\cal C$ at the boundary
between $u$ and $v$ coincides with the crossing sequence of ${\cal C}'$
at the boundary between $u'$ and $v'$. Thus, there is a computation
$\widehat{\cal C}$ on the input $u'v$ which exhibits the same crossing
sequence $c$ at the boundary between $u'$ and $v$. Furthermore,
$\widehat{\cal C}$ behaves as ${\cal C}'$ on the prefix $u'$ and as
$\cal C$ on the suffix $v$. 
Note that if the length of $c$ is odd, then 
$u'v$ is accepted or rejected by $\widehat{\cal C}$ if and only
$uv$ if it is, respectively, accepted or rejected by $\cal C$.
In a similar way, if the length of $c$ is even, then 
$u'v$ is accepted or rejected by $\widehat{\cal C}$ if and only
$u'v'$  if it is, respectively, accepted or rejected by ${\cal C}'$.
Because $\widehat{\cal C}$ is obtained by pasting together pieces
of $\cal C$ with pieces of ${\cal C}'$, according to a ``cut-and-paste''
of the inputs, we will call this method \emph{cut-and-paste}.

\smallbreak

We consider also the \emph{standard Turing machine} model,
having a finite state control, a two-way read-only input tape 
and one separate semi-infinite worktape (see, e.g.,~\cite{HU79}).

\smallbreak

For each computation $\cal C$ of a deterministic or nondeterministic, 
one-tape off-line or standard Turing machine $M$, we consider the
following resources:
\begin{itemize}
  \item The \emph{time}, denoted as $t(\cal C)$, is the number 
  of moves in the computation $\cal C$.
  \item The \emph{space}, denoted as $s(\cal C)$, is the number 
  of cells used in the computation $\cal C$. In the case of standard machines, 
  the space is measured by keeping into account only the worktape. 
  (For one-tape off-line machines we do not consider this resource in the paper.)
  \item The \emph{length of the crossing sequences}, denoted
  as $c(\cal C)$, for a one-tape off-line machine is the number of the 
  states in the longest crossing sequence used by $\cal C$. 
  (It is possible to give a similar measure even for standard machines 
  see, e.g.,~\cite{BMP94}, but we will do not make use of it in
  the paper.)
\end{itemize}
Given an input string $x$, we want to consider how much of
the above resources is used by the machine $M$, having $x$ as input. 
In the case of deterministic machines, there is only
one computation associated with each string $x$. Hence the use of 
each resource could be trivially defined referring to 
such a computation.
On the other hand, in the case of nondeterministic machines, 
we can have several computations on a same string. 
This leads to several measures. We now present and briefly discuss
those ones considered in the paper.

We say that machine $M$ uses $r(x)$ of a resource $r\in\{t,s,c\}$
(time, space, length of crossing sequences, resp.), 
on an input $x$ if and only if 
\begin{itemize}
\item \emph{\strong\ measure:}
  \[r(x)=\max\{r({\cal C})\mid\mbox{ ${\cal C}$ is 
  a computation on $x$}\}\]
\item \emph{\accept\ measure:}
  \[r(x)=\left\{
         \begin{array}{ll}
         \max\{r({\cal C})\mid\mbox{ ${\cal C}$ is
         an accepting computation on $x$}\}&\mbox{if $x\in L$}\\
         0 &\mbox{otherwise}
         \end{array}
        \right. \]\item \emph{\weak\ measure:}
  \[r(x)=\left\{
         \begin{array}{ll}
         \min\{r({\cal C})\mid\mbox{ ${\cal C}$ is
         an accepting computation on $x$}\}&\mbox{if $x\in L$}\\
         0 &\mbox{otherwise}
         \end{array}
        \right. \]
\end{itemize}
The \emph{\weak} measure corresponds to an optimistic view related to the
idea of nondeterminism: a nondeterministic machine, besides choosing
an accepting computation, if any, is able to chose that one of minimal cost.
On the opposite side, the \emph{\strong} measure keeps into account the 
costs of all possible computations. Between these two measures, the \emph{\accept}
one keeps into account the costs of \emph{all} accepting computations.
(For rejected inputs, for technical reasons, it is suitable to
set the \accept\ and the \weak\ measure to $0$). 
These notions have been proved to be different,
for example in the context of space bounded computations~\cite{Me08}.

As usual, we will mainly define complexities with respect to
input lengths.
This is done by considering the worst case among all
possible inputs of the same length. Hence, under the \strong,
\accept, and \weak\ measures, for $r\in\{t,s,c\}$, we define
\[
r(n)=\max\{r(x)\mid x\in\Sigma^*, |x|=n\}.
\]

All the logarithms are in bases $2$.
We assume that the reader is familiar with the asymptotic
notations, in particular with \emph{big-Oh} and 
\emph{little-Oh} (denoted, respectively, as $O(\cdot)$ and $o(\cdot)$),
and with basic techniques concerning Turing machines, in particular
with the use of tape tracks.

\section{Simple bounds}

In this section, we present some simple lower bounds for the recognition
of nonregular languages. Probably they are already known as ``folklore'',
but we include them for the sake of completeness and also because some
of the arguments used to prove them will be interesting later in the paper.
We formulate the bounds for the \weak\ measure. As a consequence, they
hold for all the measures considered in the paper.

First of all, considering crossing sequences, we prove the following:

\begin{theorem}\label{th:cs}
  If $L$ is accepted by a nondeterministic one-tape off-line Turing machine 
  $M$ with $c(n)=O(1)$, under the \weak\ measure, then $L$ is regular.
\end{theorem}
\begin{proof}
  Let $M$ be a nondeterministic one-tape off-line Turing machine
  accepting $L$ and $c$ a costant such that, for each $x\in L$,
  $M$ has an accepting computation on $x$ using only crossing sequences
  of length bounded by $c$.
  Without loss of generality, we can suppose that when $M$ accepts,
  its head is positioned on the leftmost square of the tape containing the
  blank symbol (namely, besides the right end of the input and of
  the tape portion used during the computation).
  
  There exists a nondeterministic finite automaton $N$, whose set 
  of states is the set of all possible sequences at most $c$ states
  of $M$, whose initial state is the sequence containing only the initial
  state of $M$, whose transition function is defined according to the 
  compatibility relation between crossing sequences described 
  in Section~\ref{sec:prel}, 
  and whose set of final states contains the sequences which, 
  at the right boundary of the input, lead to the acceptance. 
  (By definition, the machine $M$ 
  during the computation can also use a portion of the tape to the 
  right of the input string. At the beginning of the computation,
  this portions always
  contains blank symbols, hence, it does not depend on the input.
  The possible behaviors of $M$ on this portion, completely depend on
  the crossing sequence at its left boundary. In particular,
  for a given machine, it is possible to determine the
  set of crossing sequences that, to the left of a blank portion
  of the tape, lead to the acceptance.)

  It can be easily observed that each accepting computation of $N$
  simulates an accepting computation of $M$. Since
  each $x\in L$ has an accepting computation using crossing
  sequences of length at most $c$, we can conclude that $N$
  accepts exactly $L$.
\end{proof}

We now consider the time. We can observe that in each computation $\cal C$
on an input $w$, ending in less than $|w|$ steps, the head
cannot reach the right end of the input. This implies that the same
computation can be performed on each string having $w$ as a prefix.
Using this observation, it is not difficult to prove that
if a language $L$ is accepted in time $t(n)=o(n)$ under the \strong\ 
measure, then $L$ is regular and, actually, it can be
accepted in constant time.
This result can be proved also for the \weak\ measure:

\begin{theorem}\label{th:time}
  Let $M$ be a nondeterministic (one-tape off-line or standard)
  Turing machine accepting a language
  $L$ in time $t(n)=o(n)$, under the \weak\ measure. Then
  $t(n)=O(1)$ and $L$ is regular.  
\end{theorem}
\begin{proof}
  Since $t(n)=o(n)$, it should exists an integer $n_0$
  such that $t(n)<n$ for each $n\geq n_0$.
  We prove that each string belonging to $L$ is accepted by a
  computation of less than $n_0$ steps, thus implying
  $t(n)=O(1)$.
  
  Given $x\in L$ with $|x|\geq n_0$, there exists a computation accepting
  $x$ in $t(x)<|x|$ steps. Hence, such a computation can read at most the 
  first $t(x)$ symbols of $x$ and, so, it will accept even the prefix
  $x_0$ of $x$ of length $t(x)$.
  Hence, $t(x_0)\leq t(x)=|x_0|$.
  
  If $t(x_0)<|x_0|$, then there exists a computation which accepts
  $x_0$ just reading a proper prefix of $x_0$. Because $x_0$ is a
  prefix of $x$, the same computation should accept also $x$.
  Since the length of such a computation
  is less than $t(x)$, this is a contradiction.
  
  This permit us to conclude that $t(x_0)=t(x)=|x_0|$.
  Because $t(n)<n$, for $n\geq n_0$, we can conclude that 
  $|x_0|<n_0$. Hence, $x$ is accepted in less than $n_0$ steps.

\smallbreak 

  We finally point out that this result
  does not depend on the machine model: it holds for 
  one-tape off-line Turing machines as well as for standard
  machines.
\end{proof}

\section{Lower bounds for the \accept\ measure}

The problem of proving lower bounds for the length of crossing 
sequences and for the running time of one-way off-line 
Turing machines accepting nonregular languages was investigated
by several authors.
Considering deterministic machines,
in the paper of Hennie~\cite{He65} a logarithmic lower bound
for the length of the crossing sequences was proved.
Furthermore, it was shown that each language accepted
in linear time is regular. A better lower bound
(of the order of $n\log n$) for the time needed to recognize
nonregular languages was
independently proved by Trakhtenbrot~\cite{Tr64}, 
Hartmanis~\cite{Ha68}, and Kobayashi~\cite{Ko85}.

For the nondeterministic case, Wagner and Wechsung~\cite{WW86} showed
that under the \weak\ measure the same does not hold. In particular,
Michel~\cite{Mi91} gave examples of nonregular languages
accepted in linear time. However, considering the
\strong\ measure, recently Tadaki, Yamakami and Lin~\cite{TYL04} 
extended the argument used by Kobayashi to nondeterministic
machines, by  proving that in such a case,
the logarithmic lower bound on the length of the crossing sequences
and the $n\log n$ lower bound for the time needed to
accept nonregular languages still hold.

In this section, we further deepen such an 
investigation by showing that
the lower bounds for the deterministic case are true even for
nondeterministic machines, if we restrict our attention 
only to accepting computations, namely if we replace the \strong\ measure
considered in~\cite{TYL04} with the \accept\ measure.
Our proof is obtained by suitable extending 
that one presented in the book of Wagner and 
Wechsung~\cite[Th.s 8.15, 8.17]{WW86} for the deterministic case.

\begin{theorem}\label{th:accept}
  Let $M$ be a nondeterministic one-tape off-line Turing machine, using
  crossing sequences of length $c(n)$ and working in time $t(n)$ under
  the \accept\ measure. Then:
  
  (1) $c(n)=o(\log n)$ implies $c(n)=O(1)$,
  
  (2) $t(n)=o(n\log n)$ implies $t(n)=O(n)$.
  
  Furthermore, $c(n)=O(1)$ if and only if $t(n)=O(n)$. In this case the
  accepted language is regular.
\end{theorem}
\begin{proof}
  As in the proof of Theorem~\ref{th:cs}, without loss of generality
  we can suppose that $M$ accepts with the head to the right of the
  portion of the tape used during the computation.

  For each integer $k\geq 1$, let $L[k]$ be the set of strings
  having an accepting computation whose longest crossing sequence consists
  of exactly $k$ states.
  
  For $L[k]\neq\emptyset$, let $w_k$ be a shortest
  word belonging to $L[k]$ and $n_k=|w_k|$. Hence, $c(n_k)\geq k$.
  Consider a computation $\cal C$ accepting $w_k$ and using a crossing
  sequence $c_0$, occurring at a tape boundary $b_0$, with
  $|c_0|=k$. 
  
  If a same crossing sequence $c$ occurs in the computation
  $\cal C$ at three boundaries $b_1,b_2,b_3$ of the input zone, with
  $1\leq b_1<b_2<b_3\leq |w_k|$, then
  $w_k$ can be decomposed as $xyzt$, where $x$ is the prefix of $w_k$
  which ends at the boundary $b_1$, $y\neq\epsilon$ and $z\neq\epsilon$ are
  the factors delimited, respectively, by the boundaries $b_1$ and $b_2$, and
  $b_2$ and $b_3$, $t$ is the suffix which starts at the boundary $b_3$.
  We can use the ``cut--and--paste'' method discussed in 
  Section~\ref{sec:prel}, combining in several ways these 
  strings and the associated computation slices. 
  In particular, if $b_0>b_2$
  then we can get an accepting computation on the string $xzt$ that still
  uses (in the part $zt$) the crossing sequence $c_0$. In a similar way,
  if $b_0\leq b_2$ then we can get an accepting computation on the string
  $xyt$ which uses $c_0$. Hence, in both cases, we get a string
  $w'$ shorter than $w_k$ and belonging to $L[k]$. This is
  a contradiction to our choice of $w_k$.
  This permit us to conclude that every crossing sequence used in the input
  zone of $w_k$ can occur at most twice.
  
  Consequently, $w_k$ has at least $\left\lfloor\frac{n_k-1}{2}\right\rfloor$
  different crossing sequences.
  If $q$ is the number of the states of $M$, this implies
  $\left\lfloor\frac{n_k-1}{2}\right\rfloor\leq q^{k+1}$, and
  hence $c(n_k)\geq k\geq\log_{q}\left\lfloor\frac{n_k-1}{2}\right\rfloor-1$.
  
  If $c(n)$ is unbounded, then $L[k]\neq\emptyset$ for infinitely many
  $k$, thus implying $c(n)\geq d\log n$, for some constant $d>0$
  and infinitely many integers $n$, i.e., $c(n)\neq o(\log n)$.
  
  \medskip
  
  We can easily observe that $c(n)=O(1)$ implies $t(n)=O(n)$.
  Hence, if $M$ does not work in linear time, then $c(n)$ is
  unbounded and, according to the previous part of the proof, there 
  exists a sequence $w_1, w_2,\ldots$ of strings with $|w_1|<|w_2|<\ldots$
  such that $w_i$ has an accepting computation ${\cal C}_i$ using at 
  least $\left\lfloor\frac{|w_i|-1}{2}\right\rfloor$
  different crossing sequences. On the other hand, the number of
  different crossing sequences of ${\cal C}_i$ of length less than
  $\frac{\log\left\lfloor\frac{|w_i|+1}{4}\right\rfloor}{\log q}$
  is at most $\left\lfloor\frac{|w_i|+1}{4}\right\rfloor$. 
  As a consequence, the accepting computation ${\cal C}_i$ has at least
  $\left\lfloor\frac{|w_i|-1}{4}\right\rfloor$ different crossing 
  sequences of length at least 
  $\frac{\log\left\lfloor\frac{|w_i|+1}{4}\right\rfloor}{\log q}$
  and, hence, it consists of at
  least $\left\lfloor\frac{|w_i|-1}{4}\right\rfloor\cdot 
  \frac{\log\left\lfloor\frac{|w_i|+1}{4}\right\rfloor}{\log q}
  \geq d|w_i|\log(|w_i|)$ steps, for some constant $d>0$. 
  This permits us to conclude that $t(n)\geq dn\log n$, for
  infinitely many integers $n$.
  Hence, $t(n)=o(n\log n)$ implies $t(n)=O(n)$ and $c(n)=O(1)$.
  
  By summarizing the implications we have proved, we can also
  conclude that $c(n)=O(1)$ if and only if $t(n)=O(n)$.
  In this case, by Theorem~\ref{th:cs}, it follows that the
  language accepted by $M$ is regular.

\end{proof}

\section{Lower bound for the \weak\ measure}

While the \strong\ measure takes into account the costs of all computations
and the \accept\ measure the costs of all accepting
computations, the \weak\ measure restricts only to accepting
computations of minimal costs. This is closely related to the notion of
nondeterminism: a machine $M$, among all possible computations, selects
the accepting one of minimal cost.

Under the \weak\ measure, the time lower bound presented in 
Theorem~\ref{th:accept} for nonregular language acceptance
does not hold. In fact, as proved by Michel~\cite{Mi91}, 
it is possible to accept in linear time some NP-complete languages.

On the other hand, as proved in Theorem~\ref{th:cs},
a constant bound on the
length of the crossing sequences implies the regularity of the
accepted language. Hence, we can ask if it is still possible to state
a lower bound on the maximal length of the crossing sequences for the
recognition of nonregular languages, under the \weak\ measure.
The next theorem gives a positive solution to this problem,
stating a $\log\log n$ lower bound (hence, smaller than the bound for the
\accept\ measure). In the next section we will show that this lower
bound is reachable, even in the case of languages defined over a one 
letter alphabet.

\begin{theorem}\label{th:weak}
  Let $M$ be a nondeterministic one-tape off-line Turing machine, using
  crossing sequences of length at most $c(n)$ under the \weak\ measure.
  If $c(n)=o(\log\log n)$ then $M$ accepts a regular language.
\end{theorem}
\begin{proof}
  Using the same technique as in the proof of Theorem~\ref{th:cs},
  we can show that for each integer $n\geq 1$ there exists a 
  nondeterministic finite automaton $N_n$, whose states are the
  sequences of at most $c(n)$ states of $M$, behaving as $M$
  on strings of length at most $n$.
  
  Hence, denoting by $L$ the language accepted by $M$, we get that 
  $L(N_n)^{\leq n}=L^{\leq n}=L(A_n)^{\leq n}$, where $A_n$
  is a deterministic automaton equivalent to $N_n$.
  
  If $q$ is the number of the states of $M$, then
  the number of the states of $N_n$ is bounded by $q^{c(n)+1}$, and
  that of $A_n$ by $2^{2^{(c(n)+1)\log q}}$. By a result of Karp~\cite{Ka67},
  if $L$ is nonregular, such a number must be at least $\frac{n+3}{2}$,
  for infinitely many $n$. Hence, we get that $c(n)\geq d\log\log n$, for
  some constant $d$, infinitely often.
\end{proof}

In the proof of Theorem~\ref{th:weak}, for each integer $n$ we simulate 
the machine $M$ with a finite automaton, which agrees with 
$M$ on strings of length at most $n$. 
Using a similar idea, we can simulate the
machine $M$ with an equivalent standard machine $M'$ that uses a one-way read-only 
input tape.
The machine $M'$ works essentially as the automaton $N_n$,
scanning the input from the left to the right and checking the compatibility between
crossing sequences. However, the machine $M'$ does not have any limitation on
the length of the crossing sequences. This is achieved using two tracks
of the worktape. The first track contains a crossing
sequence at the left boundary of the current input square. The machine guesses
on the second track another crossing sequence and verifies whether or not
it is compatible with that on the first track, with respect to the symbol
under the input head. If the outcome of this test is negative then
the machine stops and reject, otherwise the machine continues the computation.
When the right end of the input tape is reached, the machine $M'$ can continue
by simulating the crossing sequences on the blank part of the tape of $M$.
Hence, we get the following:

\begin{theorem}\label{th:simulation}
  Each (deterministic or nondeterministic) one-tape off-line Turing machine 
  $M$ can be simulated by an equivalent one-way nondeterministic standard
  Turing machine $M'$ such that:
  \begin{itemize}
  \item If $M$ uses crossing sequences of length at most $c(n)$
    under the \weak\ measure, then $M'$ works in space $c(n)$ under the
    \weak\ measure.
  \item If $M$ uses crossing sequences of length at most $c(n)$ under
    the \strong\ or \accept\ measures, then $M'$ works in space 
    $c(n)$ under the \accept\ measure.
  \end{itemize}
\end{theorem}
\begin{proof}
  From the above outlined construction, we can observe that to each accepting 
  computation of $M$ using crossing sequences of length at most 
  $k$ corresponds an accepting computation of $M'$ using
  space $O(k)$. Hence, the two statements of the theorem easily 
  follows.\footnote{We observe that, due to the guessing process 
  implemented by $M'$,
  nonaccepting computations of $M'$ can use a huge amount
  of space, which is not related to the length of the crossing sequences 
  actually used in the computation of $M$.
  For this reason the theorem does not give any information on the space
  used by nonaccepting computations of $M'$.}
\end{proof}

Space lower bound for Turing machines accepting nonregular languages have been 
extensively investigated in the literature.
In particular, in the case of one-way nondeterministic machines, 
it is known that
in order to accept a nonregular language the space used 
by the machine must grow at least as $\log\log n$ under the 
\weak\ measure~\cite{Al85} and as $\log n$ under the
\accept\ measure~\cite{Me08}. Combining these space lower bounds 
with the result given
in Theorem~\ref{th:simulation}, we get an alternative proof of the lower 
bounds on the length of the crossing sequences presented in 
Theorem~\ref{th:accept} and in Theorem~\ref{th:weak}.

\section{Fast recognition of unary languages}

In this section we present some interesting examples of nonregular
languages accepted by one-tape off-line Turing machines,
using small amounts of time.
All the languages we consider are defined over a one letter alphabet
and have been previously presented in the literature because they have
small space complexities on standard machines.

\medskip

Before introducing the examples, we briefly recall a technique
presented in~\cite[proof of Th.\ 2.2]{Mi91} which will be extensively
used in our recognition algorithms.
This technique is useful to count input factors of length $k$, using 
$O(k\log k)$ moves, on a one-tape off-line machine:
\begin{itemize}
\item Three tape tracks are used: track 1, which is left unchanged,
  contains the input string; track 2 and track 3 are used to
  count.
\item The computation starts with the head scanning the first
  symbol of the input that must be counted, and $0$ written on
  track 2 (in the same cell scanned by the head). 
  At each step\footnote{With the term ``step'', here we intend
  the set of operations performed by the machine to count one
  input position.} the counter on track 2 is incremented.
  However, it is also shifted one position to the right
  in such a way that, after $j$ steps, the representation of
  the number $j$ on the track 2 is not ``too far'' from
  the counted position $j$. More precisely: 
\item After counting $j\leq k$ input symbols, $j$ must be written
  in a base $d$ on track 2, starting from the tape square under
  the $j$th position.
\item To increment from position $j$ to position $j+1$, the counter
  is copied and incremented, \emph{one position to the right}
  on track 3. (Then, it is copied back on track 2 for the next step,
  or the roles of tracks 2 and 3 are switched.)
\end{itemize}
The increment step from $j$ to $j+1$ uses at most $c_1\log_dj+1$ moves,
for some constant $c_1$. Since, during this process, the counter 
is incremented from $0$ to $k$, the total number of moves is bounded by 
$k(c_1\log_dk+1)\leq\frac{c_2}{\log d}k\log k=O(k\log k)$ for sufficiently 
large $k$. Hence, a factor of length $k$ of the input can be
counted in time $O(k\log k)$, by a deterministic procedure.
It can be also observed that, by suitably choosing the base $d$,
the time can be bounded by $\epsilon k\log k$, for any given $\epsilon>0$.

\medskip

We are now ready to present our first example.
We denote by $p_1,p_2,\ldots$ the sequence of prime numbers in increasing order.
We consider the language
\begin{eqnarray*}
  L_0=\{ a^n &|& \exists t\geq 1 \mbox{ s.t. $n$ is divisible by
                                       $p_1,p_2,\ldots,p_t$}\\
             & & \mbox{but not by $p_1^2,p_2^2,\ldots,p_t^2$ and
                       $p_{t+1}$}\}.
\end{eqnarray*}
Hartmanis and Berman~\cite{HB76} proved that $L_0$ is a nonregular
language accepted in $O(\log\log n)$ space by a deterministic standard
machine under the \strong\ measure. Hence, $L_0$ matches the space
lower bound for this kind of machines. We now show that this language
matches also the time lower bound proved in Theorem~\ref{th:accept}.
More precisely:

\begin{theorem}\label{th:L0}
  $L_0$ is accepted by a deterministic one-tape off-line 
  Turing machine in time $O(n\log n)$, under the \strong\ measure.
\end{theorem}
\begin{proof}
  In order to verify membership to $L_0$, the following
  immediate algorithm can be used:
\begin{tabbing}
input $a^n$\\
$i\leftarrow 1$\\
{\bf whi}\={\bf le} $p_i$ divides $n$ {\bf and} $p_i^2$ does not divide $n$ {\bf do}\\
\>$i\leftarrow i+1$\\
{\bf if }$p_i$ does not divide $n$ \= {\bf then }\= accept\\
\>{\bf else }\>reject\\
\end{tabbing}
Now, we explain how the steps of this algorithm can be implemented
in the machine model we are considering and we evaluate the time 
used by such an implementation:
\begin{itemize}
\item \emph{Testing divisibility} of $n$ by an integer $k$:\\
  Given $k$ written on a track $T$ of the tape, we can adapt the above
  described procedure, in order to count the input length modulo $k$.
  The procedure is modified in such a way that at each step, the value
  of the counter is compared with the original value of $k$, resetting
  the counter when it reaches $k$. In order to avoid the use of extra moves,
  at each step, when the counter is shifted and incremented, 
  also the representation of $k$ on the track $T$ is moved
  one position to the right.
  In this way, when the right side of the input is reached,
  the value contained in the counter will be $n\bmod k$. Hence, the
  machine can finally verify whether or not such a value is $0$.
  
  The time used to count a factor of length $k$ is $O(k\log k)$. Because
  this is reapeted $\left\lceil\frac{n}{k}\right\rceil$ times, the overall
  time used to test the divisibility is $O(n\log k)$.

\item \emph{Testing divisibility} of $n$ by $k^2$, for a given integer $k$:\\
  This can be done, while checking the divisibility of $n$ by $k$,
  just introducing another counter modulo $k$ 
  (making use of extra tracks).
  The new counter is shifted each time an input position is counted (like
  the first counter used to test the divisibility by $k$). However, it
  is incremented only when the first counter is reset, i.e., each
  time a block of $k$ input symbols has been scanned.
  
  We point out that this test can be performed in the above time bound.
  
\item \emph{Computing the sequence of primes $p_1,p_2,\ldots$}\\
  This is done by using the Eratostene sieve algorithm, implemented
  as a ``coroutine'' of the main program.
  A special track, called $P$, is used to designate the positions of 
  non prime numbers, i.e., the cell in position $j$ of $P$ will be
  marked with a special symbol $X$ after discovering that $j$ is not
  a prime number. At the beginning of the computation, all the cells
  are unmarked. When the machine tests the divisibility of $n$ by a prime
  $p_i$, using the above described procedure,
  it can also mark on the track $P$ the positions corresponding
  to multiples of $p_i$ (different from $p_i$). Hence, also the 
  generation of the primes
  can be done within the previous time bound.
  
\item \emph{Preparing the next iteration: $i\leftarrow i+1$}\\
  Actually, we do not explicitly need  the value of $i$, but
  we have to switch from a prime $p_i$ to the next prime $p_{i+1}$.
  To this aim, on the track $P$, the first unmarked position 
  to the right of the position $p_i$  must be searched.
  We remind that the prime $p_i$ was kept on the track $T$ (it has
  been shifted to the right side, to test the divisibility of $n$
  by $p_i$, but after that it can be moved back within
  $O(n\log p_i)$ steps).
  Starting from the left side of the tape, the position $p_i$ is
  reached (counting until $p_i$). Hence, the positions to its right side
  are scanned (by continuing to count), until to reach the first
  square which on the track $P$ is unmarked. The position of such a
  square (available in the counter) gives the next prime $p_{i+1}$. 
  Finally, the representation of $p_{i+1}$ is shifted back, in order to
  have it at the beginning of track $T$, to be used in the
  next iteration.
  
  The number of steps used to get $p_{i+1}$ from $p_i$ by this
  procedure is $O(p_{i+1}\log p_{i+1})$, which is bounded by
  $O(n\log p_i)$ because the primes we have to consider do not
  exceed $n$ and $p_{i+1}<2p_i$~\cite{HW79}. 
    
  Hence, the total number of steps used to prepare the iteration $i+1$
  is $O(n\log p_i)$.
\end{itemize}
We can now evaluate the overall time used by this procedure. 
Given an integer $n$, let $x$ be the first prime not dividing $n$.
The above described algorithm
accepts $a^n$, if it belongs to $L_0$, ending with $p_i=x$. On the other
hand, if
$a^n\notin L_0$ then the algorithm rejects with $p_i<x$.
In both cases, the time used is bounded by:
\[\sum_{p_i\leq x}cn\log p_i=cn\sum_{p_i\leq x}\log p_i,\]
for some constant $c$. 
Since the sum on the (natural) logarithms of all primes not
exceeding an integer $x$ is asymptotic to $x$ and the smaller prime
not dividing an integer $n$ is $O(\log n)$~\cite{HW79,HB76},
the value of the last sum is $O(\log n)$
Hence, we can finally conclude that the total running time in $O(n\log n)$.
\end{proof}

We give a further example of language recognized in time $O(n\log n)$ by
a determistic machine under the \strong\ measure. Such a language was
presented by Alt and Mehlhorn~\cite{AM75}, that showed that it is accepted
by a deterministic standard machine in space $O(\log\log n)$ (minimal amount
of space needed to recognize nonregular languages). That result
was improved in~\cite{BMP94}, where it has been shown how to recognize the
same language with a deterministic standard
machine that uses the same amount of space and
$O(\frac{\log n}{\log\log n})$ input head reversals (minimum
amount needed for the recognition of nonregular languages by machines
using the minimum amount of space $O(\log\log n)$).

Given a nonnegative integer $n$, let us denote by $q(n)$ the smallest
integer non dividing $n$. We consider the language
\[
L_{AM}=\{a^n\mid q(n) \mbox{ is a power of 2}\}
\]

\begin{theorem}\label{th:am}
  $L_{AM}$ is accepted by a deterministic one-tape off-line 
  Turing machine in time $O(n\log n)$, under the \strong\ measure.
\end{theorem}
\begin{proof}
  In~\cite{BMP94}, it is proved that for each integer $n$, the number $q(n)$
  is a power of a prime number.
  As a consequence, an algorithm was described that divides $n$ by all
  prime powers, considered in the increasing order, until to find $q(n)$.
  
  We can implement such an algorithm by a one-tape off-line Turing machine,
  along the same ideas used in the proof of Theorem~\ref{th:L0}.
  The main difference is that in this case we have to test divisibility
  of the input length not only for prime numbers, but also for their powers.
  To this aim, we modify the routine implementing the Eratostene sieve as
  we now explain. We use two different symbols $X$ and $Y$ to mark the 
  positions on the track $P$: when we have to mark the square at position $k$,
  because we discovered that $k$ is a multiple of a prime number, we first
  check the content of the square. If the square is unmarked, then we mark 
  it with the symbol $X$ (this means that just one prime divisor 
  of $k$ has been found), otherwise, we mark it with $Y$ 
  (this means that at least two different primes dividing $k$ have been found).
  In this way, up to a certain value, the prime powers corresponds to 
  the positions that in the track $P$ are unmarked or are marked with $X$.
  
  Finally, when $q(n)$ has been computed, the machine must verify
  whether or not it is a power of $2$. This can be trivially done if the
  counters used in the computation are represented in base $2$.
  
  We can now estimate the time used by this implementation of the algorithm.
  For each prime power $k$, $O(n\log k)$ steps are 
  used to check the divisibility
  of $n$ by $k$. Furthermore, the number
  of prime powers not exceeding an integer $m$ is 
  $O(\frac{m}{\log m})$~\cite{BMP94}, and $q(n)=O(\log n)$~\cite{AM75}.
  Hence, for some constant $c$, the time is at most
  \[
  \sum_{\stackrel{k\leq q(n)}{\mbox{\tiny $k$ prime power}}}\!\!\!\!\!\!\!\!cn\log k=
  cn\log q(n)\cdot\frac{q(n)}{\log q(n)} = cn \cdot q(n) = O(n\log n).
  \]
\end{proof}

The languages $L_0$ and $L_{AM}$ are interesting examples of languages
accepted with a minimal amount of resources. 
Besides to be examples of languages accepted using a small amount
of space by standard machines (under the \strong\ and \accept\ 
measures)~\cite{Me08}, they use also
a minimal amount of time (under the same measures) on one-tape
off-line Turing machines, as proved in Theorems~\ref{th:L0} and~\ref{th:am},
showing in this way that the lower bound
stated in Theorem~\ref{th:accept} is tight even in the case of
unary languages.

The complement of $L_{AM}$ seems to be even more interesting. 
In fact, as shown~\cite{Me08}, under the \weak\ measure
it can be accepted in space $O(\log\log n)$ even by a standard \emph{one-way}
machine, proving the optimality of the corresponding space
lower bound in the unary case (a similar result it is not known for $L_0$
and $L_{AM}$). Concerning one-tape off-line Turing machines
accepting the complement of $L_{AM}$, we can prove the following:

\begin{theorem}\label{th:amc}
  The complement of $L_{AM}$ is accepted by a nondeterminstic one-tape off-line  
  Turing machine in time $O(n\log\log n)$, using crossing sequences of
  length $O(\log\log n)$, under the \weak\ measure.
\end{theorem}
\begin{proof}
  It can be observed that for each integer $n\geq 1$, $q(n)$ is not
  a power of 2 if and only if there are two positive integer $s$ and $t$
  such that $2^s<t<2^{s+1}$, $n\bmod 2^s=0$, and $n\bmod t\neq 0$.
  Using such a property, in~\cite{Me08} the following nondeterministic
  algorithm for the recognition of the complement $L_{AM}$
  was presented:
\begin{tabbing}
input $a^n$\\
guess an integer $s$, $s>1$\\
guess an integer $t$, $2^s<t<2^{s+1}$\\
{\bf if} \= $n\bmod 2^s = 0$ {\bf and} $n\bmod t\neq 0$
\={\bf then} \= accept\\
\>\>{\bf else} \> reject
\end{tabbing}
The above algorithm can be implemented as follows: one track
of the tape is used to guess a power of $2$ in binary notation,
while another track is used to guess the integer $t$.
By making use of the technique described in the proof of 
Theorem~\ref{th:L0}, the machine can check in time $O(n\log t)$ 
the divisibility of $n$ by $2^s$ and by $t$, and, hence,
accept or reject the input.
If $a^n\notin L_{AM}$ then there is an accepting computation such that $t=q(n)$.
Furthermore, $q(n)=O(\log n)$~\cite{AM75}.
This permit us to conclude that the time used by the machine, under the
\weak\ measure, is $O(n\log\log n)$.
\end{proof}

Theorem~\ref{th:amc} proves the optimality of the lower bound
given in Theorem~\ref{th:weak} for the length of the crossing
sequences in the case of the \weak\ measure. Concerning the time, we
recall that in~\cite{Mi91}, the existence of NP-complete 
languages accepted in linear time, under the \weak\ measure, have been proved.
Such languages are obtained using a padding
technique which relies on the use of an input alphabet with more than one
symbol. We conjecture that, in the unary case, the recognition of 
unary nonregular
languages by nondeterministic one-tape off-line Turing machines requires, 
under the \weak\ measure, 
more than linear time. More precisely, in the light of Theorem~\ref{th:amc},
we strongly believe that each unary language accepted in time $o(n\log\log n)$
under the \weak\ measure is regular.

\section{Conclusion}

In the paper the lower bounds obtained by~Trakhtenbrot~\cite{Tr64}, 
Hennie~\cite{He65}, and Hartmanis~\cite{Ha68} in the case of 
deterministic machines have been mentioned several times. 
Those results refer to the
\strong\ measure, which is usually considered
in the case of deterministic computations~\cite{WW86}. 
However, even in the deterministic case, it can be interesting
to know what happens by considering only the costs
of accepting computations. Because a deterministic machine
cannot have more than one computation on each input string,
it turns out that, in the deterministic case, the \weak\ and the \accept\ 
measures coincide.
Hence, the lower bounds proved in Theorem~\ref{th:accept}
in the case of nondeterministic machines under the
\accept\ measure hold even for deterministic machines
under the \weak\ measure. Because they coincide with the
lower bound for the \strong\ measure, which is known to be 
optimal, they cannot be increased, i.e., they are also optimal.

In Table~\ref{tb:time} the time lower bounds for the recognition
of nonregular languages by one-tape off-line Turing machines
are summarized. The table should be
read as follows: a row $r$ denotes a type of machine while
a column $c$ a measure. If the element at the position $(r,c)$ 
of the table is the function $f(n)$, then $t(n)\notin o(f(n))$
for each one-tape off-line 
Turing machine of type $r$ recognize a nonregular language
in time $t(n)$ under the measure corresponding to column $c$. 
Table~\ref{tb:cs} gives a similar overview
for the lower bounds on the maximal length $c(n)$ of the crossing sequences,
used in nonregular language recognition.

The numbers in the tables refer to the following list, where a short
justification of each result is given:
\begin{enumerate}
\item Proved by Trakhtenbrot~\cite{Tr64} and Hartmanis~\cite{Ha68}. 
   Hennie~\cite{He65} proved the same
  lower bound for $c(n)$ and a smaller lower bound for the time.
\item Proved by Tadaki, Yamakami, and Lin~\cite{TYL04}.
\item Theorem~\ref{th:accept}.
\item Consequence of 3.
\item Consequence of 4.
\item Theorem~\ref{th:time} and Theorem~\ref{th:weak}.
\end{enumerate}
\begin{table}[hbt]\footnotesize
\begin{center}
\begin{tabular}{l@{\,\,}r@{}}
\begin{tabular}[b]{|l|}\multicolumn{1}{c}{}\\ \hline
~Deterministic machines~\\ \hline
~Nondeterministic machines~\\ \hline
\end{tabular}&
\begin{tabular}[b]{|c|c|c|@{}}
\hline {\strong}  & {\accept} & {\weak}\\ \hline\hline
$~~n\log n~~^1$ & $~~n\log n~~^4$ & $~~n\log n~~^5$ \\ \hline
$~~n\log n~~^2$ & $~~n\log n~~^3$ & $~~\hfill n~\hfill^6$\\ \hline
\end{tabular}
\end{tabular}
\caption{Lower bounds for $t(n)$}
\label{tb:time}
\end{center}
\end{table}
\begin{table}[hbt]\footnotesize
\begin{center}
\begin{tabular}{l@{\,\,}r@{}}
\begin{tabular}[b]{|l|}\multicolumn{1}{c}{}\\ \hline
~Deterministic machines~\\ \hline
~Nondeterministic machines~\\ \hline
\end{tabular}&
\begin{tabular}[b]{|c|c|c|@{}}
\hline {\strong}  & {\accept} & {\weak}\\ \hline\hline
$~~\log n~~^1$ & $~~\log n~~^4$ & $~~\hfill\log n~\hfill^5$ \\ \hline
$~~\log n~~^2$ & $~~\log n~~^3$ & $~~\log\log n~^6$\\ \hline
\end{tabular}
\end{tabular}
\caption{Lower bounds for $c(n)$}
\label{tb:cs}
\end{center}
\end{table}

I like to complete the paper with some final remarks,
with the hope to stimulate the reader, and myself, for future
researches.
As in the classical studies in complexity, in the paper I
made a large use of asymptotic notations, tape tracks (that,
in real implementations, mean large alphabets), and so on.
I believe that researches about computing with restricted
resources deserve closer investigations and refinements. 
It should be useful, for instance, to study what is hidden in the
big-Oh's and how other parameters (for example the cardinality
of the alphabet used by the machine) influence the complexity.
A recent example of interesting research in this direction is~\cite{HMT05}.
Furthermore, I think that the relationships with the descriptional 
complexity should be investigated, for example studying how the succinctness
of the description of a machine can affect its computation
time or the used space.

\section*{Acknowledgment}
I would like to thank the anonymous referee
for his valuable comments and suggestions.

\end{document}